\documentclass[aip,preprint]{revtex4-2}
\usepackage[cp1251]{inputenc}
\usepackage[T2A]{fontenc}
\usepackage[english]{babel}
\usepackage{amssymb,latexsym,amsmath,amscd}
\usepackage{graphicx,color,framed}
\usepackage{bm}
\usepackage{tikz}
\usepackage{xcolor}
\usepackage{siunitx}
\usepackage{empheq}
\usepackage{xr}
\usepackage{microtype}
\usepackage{amsfonts}
\usepackage[normalem]{ulem}
\graphicspath{{figures/}}

\makeatletter
\newcommand*{\addFileDependency}[1]{
  \typeout{(#1)}
  \@addtofilelist{#1}
  \IfFileExists{#1}{}{\typeout{No file #1.}}
}
\makeatother

\begin{document}
\affiliation{Laboratory of Computational Physics, HSE University, Tallinskaya st. 34, 123458 Moscow, Russia}
\author{\firstname{Yury A.} \surname{Budkov}}
\affiliation{Laboratory of Computational Physics, HSE University, Tallinskaya st. 34, 123458 Moscow, Russia}
\affiliation{Laboratory of Multiscale Modeling of Molecular Systems, G.A. Krestov Institute of Solution Chemistry of the Russian Academy of Sciences, 153045, Akademicheskaya st. 1, Ivanovo, Russia}
\email[]{ybudkov@hse.ru}
\title{Variational field theory of macroscopic forces in Coulomb fluids}
\author{\firstname{Petr E.} \surname{Brandyshev}}
\affiliation{Laboratory of Computational Physics, HSE University, Tallinskaya st. 34, 123458 Moscow, Russia}
\begin{abstract}
Based on the variational field theory framework, we extend our previous mean-field formalism, taking into account the electrostatic correlations of the ions. We employ a general covariant approach and derive a total stress tensor that considers the electrostatic correlations of ions. This is accomplished through an additional term that depends on the autocorrelation function of local electric field fluctuations. Utilizing the derived total stress tensor and applying the mechanical equilibrium condition, we establish a general expression for the disjoining pressure of the Coulomb fluids, confined in a pore with a slit-like geometry. Using this equation, we derive an asymptotic expression for the disjoining pressure in a slit-like pore with non-electrified conductive walls. Present theory is the basis for future modeling of the mechanical stresses that occur in electrode pores with conductive charged walls, immersed in liquid phase electrolytes beyond the mean-field theory.
\end{abstract}

\maketitle

\section{Introduction}
Coulomb fluids, such as plasma, electrolyte solutions, molten salts, and room-temperature ionic liquids, have garnered significant attention from researchers and chemical engineers in recent times. This interest is primarily driven by the utilization of Coulomb fluids in various applications, ranging from lipid and ion exchange membranes to biomacromolecules, colloids, batteries, fuel cells, supercapacitors, {\sl etc.}. In all of these applications, Coulomb fluids experience interactions with charged surfaces or confinement within nanocapillaries, leading to significant spatial inhomogeneity. This inhomogeneity of the Coulomb fluid leads to a violation of local electrical neutrality, which requires the numerical solution of the self-consistent field equations for the electrostatic potential, accompanied by appropriate boundary conditions~\cite{blossey2023poisson,naji2013perspective,budkov2022modified,budkov2021electric,avni2020charge}. The equations commonly used for this purpose are the classical Poisson-Boltzmann equation (PB) and its modified forms, which are often referred to as modified PB equations~\cite{budkov2022modified,blossey2017structural,budkov2021theory}.

The modified PB equations derived so far incorporate the unique characteristics of individual ionic species and consider the specific chemical properties of the surfaces with which they interact. In this case it is necessary to especially emphasize the consideration of the non-zero excluded volume of ions (steric interactions)~\cite{kralj1996simple,borukhov1997steric,maggs2016general,kornyshev2007double}, their specific and structural interactions~\cite{goodwin2017mean_,budkov2018theory,blossey2017structural,vasileva2023theory,budkov2023macroscopic}, polarizability/quadrupolarizability and permanent dipole moment of ionic species and solvent (co-solvent) molecules~\cite{frydel2011polarizable,budkov2020two,budkov2021theory,budkov2015modified,budkov2016theory_,budkov2018theory,coalson1996statistical,abrashkin2007dipolar,iglivc2010excluded,buyukdagli2013microscopic,budkov2023dielectric,slavchov2014quadrupole,slavchov2014quadrupole_,budkov2019statistical,avni2020charge}, surface charge regulaton~\cite{podgornik2018general_}, specific ion-electrode interactions (adsorption or depletion)~\cite{budkov2018theory,uematsu2018effects}, and electrostatic correlations~\cite{netz2000beyond,de2020continuum_,bazant2011double_,wang2010fluctuation_,wang2013effects,wang2014continuous,wang2015theoretical,agrawal2022electrostatic_,buyukdagli2020schwinger}.

Recently, Budkov and Kolesnikov~\cite{budkov2022modified} introduced a first-principle approach that enables the derivation of the modified PB equations. These equations are derived from a unified standpoint, treating them as the Euler-Lagrange equations of the grand thermodynamic potential (GTP). This functional is defined in terms of the electrostatic potential and considers factors such as the excluded volume of ions and solvent molecules as well as their static and orientation polarizabilities. Additionally, basing on the Noether's theorem the authors derived stress tensors that are consistent with the modified PB equations. This theoretical framework offers the opportunity to calculate the macroscopic forces acting on conductive or dielectric bodies immersed in the Coulomb fluid, including electrodes, colloid particles, membranes, micelles, and dust particles. While the formulation of a systematic mean-field theory for Coulomb fluids has been successful, there still remains a lack of clarity regarding the incorporation of the electrostatic correlations~\cite{naji2013perspective} into the total stress tensor. In other words, it becomes necessary to expand our formalism to encompass the electrostatic correlations of the ions. However, when considering electrostatic correlations, the resulting functionals consistently become nonlocal~\cite{netz2000beyond,lau2008fluctuation,wang2010fluctuation,wang2013effects,wang2014continuous,wang2015theoretical,wang2016inhomogeneous,buyukdagli2020schwinger}, which presents challenges for the application of Noether's theorem, originally formulated for local functionals~\cite{noether1971invariant}.

An alternative approach to Noether's theorem for this case could be the general covariant approach recently proposed by us in~\cite{brandyshev2023noether}. Our approach relies on Noether's second theorem, enabling us to derive the symmetric stress tensor for an arbitrary (nonlocal) model of the inhomogeneous liquid as a functional derivative of a GTP with respect to the metric tensor. We have applied this approach to several phenomenological models~\cite{blossey2017structural,bazant2011double_} of inhomogeneous Coulomb fluids considering electrostatic correlations of ions or short-range correlations related to packing effects.

{The variational field theory treating the electrostatic potential and the Green function as variational variables using the Gibbs-Bogoliubov inequality is largely based on the work by Wang~\cite{wang2010fluctuation}. We will extend Wang's variational approach to the arbitrary reference system of the ionic fluid. This approach, as in the case of original Wang's approach, will allow us to derive the GTP of an inhomogeneous Coulomb fluid as a functional of the electrostatic potential and a trial electrostatic Green function for an arbitrary reference fluid system. Through this derivation we will obtain self-consistent field equations for the electrostatic potential and the Green function, serving as the Euler-Lagrange equations for this functional. Thus, we will extend our previous mean-field formalism~\cite{budkov2022modified} to include the electrostatic correlations of ions. A main new result of this paper is the derivation of a total stress tensor that incorporates these electrostatic correlations, through an additional term that depends on the autocorrelation function of the local electric field fluctuations, using our general covariant approach~\cite{brandyshev_budkov_2023}. Using this derived total stress tensor, we will then obtain the general expression for the disjoining pressure in an ionic fluid confined within a slit-like pore with conductive walls and determine its asymptotic behavior in the case of non-electrified walls for very large pore thickness.}

\section{Variational field theory of Coulomb gas}
In this section, we will formulate the variational field theory for the Coulomb gas, specifically when there are no steric interactions between ions. This presentation will draw heavily from research by Wang~\cite{wang2010fluctuation} mentioned in Introduction and thus serves a pedagogical purpose, while also being a reference for the subsequent sections.

Let us consider the case of Coulomb gas with the following total potential energy of interactions
\begin{equation}
\label{U}
U=\frac{1}{2}\left(\hat{\rho} G_0\hat{\rho}\right)+\sum\limits_{\alpha}(\hat{n}_{\alpha}u_{\alpha}),
\end{equation}
where we have introduced the following short-hand notations
\begin{equation}
\left(\hat{\rho} G_0\hat{\rho}\right)=\int d\bold{r}\int d\bold{r}' \hat{\rho}(\bold{r}) G_0(\bold{r},\bold{r}')\hat{\rho}(\bold{r}')
\end{equation}
and 
\begin{equation}
(\hat{n}_{\alpha}u_{\alpha})=\int d\bold{r} \hat{n}_{\alpha}(\bold{r})u_{\alpha}(\bold{r})
\end{equation}
with
\begin{equation}
\hat{\rho}(\bold{r})=\sum\limits_{\alpha}\sum\limits_{j_{\alpha}=1}^{N_{\alpha}}\varrho_{\alpha}(\bold{r}-\bold{r}_{j_{\alpha}}),    
\end{equation}
being the microscopic density of the ions with the internal charge densities $\varrho_{\alpha}(\bold{r})$ satisfying the normalization condition
\begin{equation}
\int d\bold{r}\varrho_{\alpha}(\bold{r})=q_{\alpha},
\end{equation}
where $q_{\alpha}$ is the total electric charge of the $\alpha$th ion.
Thus, we account for the electrostatic self-interaction of the ions with nonlocal charge distributions. {Note that the concept of the internal charge densities was introduced by ref.~\cite{wang2010fluctuation} to regulate the singularity in the field theory and to yield the correct Born energy.}

The second term on the right-hand side of the eq. (\ref{U}) is the total potential energy of interaction of the ions with external fields with potentials $u_{\alpha}(\bold{r})$; the microscopic concentrations of the ions are
\begin{equation}
\hat{n}_{\alpha}(\bold{r})=\sum\limits_{j_{\alpha}=1}^{N_{\alpha}} \delta(\bold{r}-\bold{r}_{j_{\alpha}}).
\end{equation}
The two-point function $G_0(\bold{r},\bold{r}')$ is the standard Green function of the Poisson equation satisfying the following equation
\begin{equation}
-\varepsilon \nabla^2 G_0(\bold{r},\bold{r}')=\delta(\bold{r}-\bold{r}'),
\end{equation}
where $\varepsilon$ is the permittivity of medium. In the case of an infinite continuous dielectric medium, Green function has the form of standard Coulomb law, i.e. $G_{0}(\bold{r},\bold{r}')=1/4\pi \varepsilon|\bold{r}-\bold{r}'|$.

Further, using the standard Hubbard-Stratonovich (HS) transformation, we can represent the grand partition function of the Coulomb gas as the following functional integral over the fluctuating electrostatic potential~\cite{naji2013perspective,blossey2023poisson,netz2000beyond,lau2008fluctuation}
\begin{equation}
\Xi = \int \frac{\mathcal{D}\varphi}{C_0}\exp\left[-\frac{\beta}{2}\left(\varphi G_0^{-1}\varphi\right)+\sum\limits_{\alpha}z_{\alpha}\int d\bold{r}e^{i\beta \varrho_{\alpha}\varphi(\bold{r})-\beta u_{\alpha}(\bold{r})}\right],
\end{equation}
where 
\begin{equation}
C_0=\int {\mathcal{D}\varphi}\exp\bigg{[}-\frac{\beta}{2}\left(\varphi G_0^{-1}\varphi\right)\bigg{]}
\end{equation}
is the normalization constant of the Gaussian measure with the following short-hand notations
\begin{equation}
\left(\varphi G_0^{-1}\varphi\right)=\int d\bold{r}\int d\bold{r}' \varphi(\bold{r}) G_0^{-1}(\bold{r},\bold{r}')\varphi(\bold{r}'),
\end{equation}
\begin{equation}
\label{empty_space_Green}
G_0^{-1}(\bold{r},\bold{r}')=-{\varepsilon}\nabla^2\delta(\bold{r}-\bold{r}'),
\end{equation}
$z_{\alpha}=\Lambda_{\alpha}^{-3}e^{\beta\mu_{\alpha}}\theta_{\alpha}$ are the fugacities of the ionic species, $\Lambda_{\alpha}$ is the thermal wavelength of the ions of $\alpha$th kind, $\mu_{\alpha}$ -- their chemical potential, $\theta_{\alpha}$ -- their internal partition function, $\beta=(k_{B}T)^{-1}$, $k_{B}$ is the Boltzmann constant, $T$ is the temperature; we have also introduced the notation
\begin{equation}
\varrho_{\alpha}\varphi(\bold{r})=\int d\bold{r}'\varrho_{\alpha}(\bold{r}-\bold{r}')\varphi(\bold{r}').
\end{equation}

Performing the shift $\varphi \rightarrow \varphi+\varphi_0$ of the integration variable and going to a trial Green function, $G_0\rightarrow G$, we obtain
\begin{multline}
\Xi = \frac{C}{C_0}\int \frac{\mathcal{D}\varphi}{C}\exp\bigg{[}-\frac{\beta}{2}\left(\varphi G^{-1}\varphi\right)-\frac{\beta}{2}\left(\varphi [G_0^{-1}-G^{-1}]\varphi\right)-\\\beta (\varphi_0 G_0^{-1}\varphi)-\frac{\beta}{2}(\varphi_0G_{0}^{-1}\varphi_0)+\sum\limits_{\alpha}z_{\alpha}\int d\bold{r}e^{i\beta \varrho_{\alpha}\varphi_0(\bold{r})+i\beta \varrho_{\alpha}\varphi(\bold{r})-\beta u_{\alpha}(\bold{r})}\bigg{]},
\end{multline}
where we introduced a notation for normalization constant of new Gaussian measure
\begin{equation}
C=\int {\mathcal{D}\varphi}\exp\bigg{[}-\frac{\beta}{2}\left(\varphi G^{-1}\varphi\right)\bigg{]}.
\end{equation}
Further,  using the Gibbs-Bogolyubov inequality 
\begin{equation}
\left<e^{X}\right>\geq e^{\left<X\right>},
\end{equation}
where 
\begin{equation}
\left<(...)\right>=\int \frac{\mathcal{D}\varphi}{C}\exp\bigg{[}-\frac{\beta}{2}\left(\varphi G^{-1}\varphi\right)\bigg{]}(...),
\end{equation}
we arrive at
\begin{equation}
\Xi \geq \exp W[G;\varphi_0],
\end{equation}
where we introduced the following auxiliary functional
\begin{multline}
W[G;\varphi_0]= \frac{1}{2}\ln \frac{Det G}{Det G_0}-\frac{1}{2}tr\left(G [G_0^{-1}-G^{-1}]\right)-\frac{\beta}{2}(\varphi_0G_{0}^{-1}\varphi_0)+\\\sum\limits_{\alpha}{z}_{\alpha}\int d\bold{r}e^{i\beta \varrho_{\alpha}\varphi_0(\bold{r})-\frac{\beta}{2}\varrho_{\alpha}G\varrho_{\alpha}-\beta u_{\alpha}(\bold{r})}
\end{multline}
and took into account that 
\begin{equation}
\left<e^{i\beta \varrho_{\alpha}\varphi(\bold{r})}\right>=e^{-\frac{\beta}{2}\varrho_{\alpha}G\varrho_{\alpha}}=A_{\alpha}(\bold{r}),
\end{equation}
\begin{equation}
\varrho_{\alpha}G\varrho_{\alpha}=\int d\bold{r}_{1}\int d\bold{r}_2\varrho_{\alpha}(\bold{r}_{1}-\bold{r})G(\bold{r}_1,\bold{r}_2)\varrho_{\alpha}(\bold{r}_{2}-\bold{r}),
\end{equation}
and that $C/C_0=Det G/ Det G_0$, where the symbol $Det$ denotes the functional determinant of the operator~\cite{zinn2002quantum_}.

The functions $\varphi_0=i\psi$ and $G(\bold{r},\bold{r}')$ are determined from the self-consistent equations 
\begin{equation}
\frac{\delta W}{\delta \varphi_0(\bold{r})}\bigg{|}_{\varphi_{0}=i\psi}=0,~\frac{\delta W}{\delta G(\bold{r},\bold{r}')}=0
\end{equation}
{which were for previously proposed in papers~\cite{wang2010fluctuation,shen2018polyelectrolyte,wang2015theoretical,wang2016inhomogeneous,lue2006variational_}.These equations guarantee the best variation estimation of the grand partition function.}

The first of them is
\begin{equation}
\nabla^2 \psi(\bold{r})=-\frac{1}{\varepsilon}\sum\limits_{\alpha}z_{\alpha} \int d\bold{r}'\varrho_{\alpha}(\bold{r}-\bold{r}') A_{\alpha}(\bold{r}')e^{-\beta \varrho_{\alpha}\psi(\bold{r}')-\beta u_{\alpha}(\bold{r}')}.
\end{equation}
To obtain the second equation, we use the identity 
\begin{equation}
\frac{1}{2}\ln\frac{Det G}{Det G_0}= \frac{1}{2}\left(tr \ln G -tr \ln G_0\right),
\end{equation}
where the trace is
\begin{equation}
tr A =\sum\limits_{n} a_{n},
\end{equation}
where $a_{n}$ are the eigenvalues of the operator $A$ in the orthonormal basis of its eigenfunctions $\psi_n(\bold{r})$, so that
\begin{equation}
a_{n}=\int d\bold{r}\int d\bold{r'} \psi_{n}(\bold{r})A(\bold{r},\bold{r}')\psi_{n}(\bold{r}'),
\end{equation}
and the kernel of the operator is
\begin{equation}
A(\bold{r},\bold{r}')=\sum\limits_{n}a_{n}\psi_{n}(\bold{r})\psi_{n}(\bold{r}'),
\end{equation}
so that 
\begin{equation}
tr A =\int d\bold{r}A(\bold{r},\bold{r}).
\end{equation}

The kernel of inverse operator is 
\begin{equation}
A^{-1}(\bold{r},\bold{r}')=\sum\limits_{n}\frac{1}{a_{n}}\psi_{n}(\bold{r})\psi_{n}(\bold{r}').
\end{equation}
Thus, we have
\begin{equation}
\frac{\delta}{\delta G(\bold{r},\bold{r}^{\prime})}tr \ln G = \frac{\delta}{\delta G(\bold{r},\bold{r}^{\prime})}\sum\limits_{n} \ln g_{n}=\sum\limits_{n}
\frac{1}{g_n}\frac{\delta g_n}{\delta G(\bold{r},\bold{r}^{\prime})},
\end{equation}
where $g_n$ are the eigenvalues of operator $G$.

Furthermore, we have
\begin{equation}
\frac{\delta g_n}{\delta G(\bold{r},\bold{r}^{\prime})}=\frac{\delta }{\delta G(\bold{r},\bold{r}^{\prime})} \int d\bold{r}_1\int d\bold{r}_2 \psi_{n}(\bold{r}_1)G(\bold{r}_1,\bold{r}_2)\psi_{n}(\bold{r}_2)= \psi_{n}(\bold{r})\psi_{n}(\bold{r}'),
\end{equation}
where we took into account that
\begin{equation}
\frac{\delta G(\bold{r}_1,\bold{r}_2)}{\delta G(\bold{r},\bold{r}')}=\delta(\bold{r}_1-\bold{r}')\delta(\bold{r}_2-\bold{r}').
\end{equation}
Then, we have
\begin{equation}
\label{deltaD}
\frac{\delta}{\delta G(\bold{r},\bold{r}^{\prime})}tr \ln G =\sum\limits_{n}
\frac{1}{g_n}\psi_{n}(\bold{r})\psi_{n}(\bold{r}')=G^{-1}(\bold{r},\bold{r}').
\end{equation}

Furthermore, we have
\begin{multline}
\frac{\delta}{\delta G(\bold{r},\bold{r}^{\prime})}tr\left(G[G_0^{-1}-G^{-1}]\right)=\frac{\delta}{\delta G(\bold{r},\bold{r}^{\prime})}tr\left(GG_0^{-1}-I\right)= \frac{\delta}{\delta G(\bold{r},\bold{r}^{\prime})}tr(GG_0^{-1})=\\\frac{\delta}{\delta G(\bold{r},\bold{r}^{\prime})}\int d\bold{r}_{1}\int d\bold{r}_2 G(\bold{r}_1,\bold{r}_2)G_0^{-1}(\bold{r}_1,\bold{r}_2)=G_0^{-1}(\bold{r},\bold{r}^{\prime})
\end{multline}
and
\begin{multline}
\frac{\delta}{\delta G(\bold{r},\bold{r}^{\prime})}\left(z_{\alpha}\int d\bold{x} A_{\alpha}(\bold{x})e^{-\beta q_{\alpha} \psi(\bold{x})-\beta u_{\alpha}(\bold{x})}\right)=-\frac{\beta z_{\alpha}}{2}\int d\bold{x}A_{\alpha}(\bold{x})\times\\ e^{-\beta q_{\alpha} \psi(\bold{x})-\beta u_{\alpha}(\bold{x})}\varrho_{\alpha}(\bold{r}-\bold{x})\varrho_{\alpha}(\bold{r}'-\bold{x}),
\end{multline}
where we have used the expression
\begin{equation}
\frac{\delta A_{\alpha}(\bold{x})}{\delta G(\bold{r},\bold{r}^{\prime})}=-\frac{\beta}{2}A_{\alpha}(\bold{x})\varrho_{\alpha}(\bold{r}-\bold{x})\varrho_{\alpha}(\bold{r}'-\bold{x}).  
\end{equation}

Therefore, the second equation takes the form 
\begin{equation}
\label{G-1_2}
G^{-1}(\bold{r},\bold{r}')-G_0^{-1}(\bold{r},\bold{r}')=\Sigma(\bold{r},\bold{r}'),
\end{equation}
where
\begin{equation}
\label{kernel}
\Sigma(\bold{r},\bold{r}')=\beta \sum \limits_{\alpha} z_{\alpha}\int d\bold{x} A_{\alpha}(\bold{x})e^{-\beta \varrho_{\alpha}\psi(\bold{x})-\beta u_{\alpha}(\bold{x})}\varrho_{\alpha}(\bold{r}-\bold{x})\varrho_{\alpha}(\bold{r}'-\bold{x}).
\end{equation}
Using the expression (\ref{empty_space_Green}) for the "empty space" inverse Green function, expression (\ref{G-1_2}) for the trial Green function $G^{-1}$ can be rewritten in the form
\begin{equation}
\label{Green_func}
\left(- \varepsilon\nabla^2+\Sigma\right)G(\bold{r},\bold{r}')=\delta(\bold{r}-\bold{r}'),
\end{equation}
where $\Sigma$ is the integral operator with the kernel (\ref{kernel}).

Thus, using the charging approach (see ref.~\cite{lue2006variational_,wang2015theoretical,wang2016inhomogeneous}), the expression for the grand thermodynamic potential (GTP) can be reduced to
\begin{multline}
\label{Omega_ion_gas_}
\Omega = -k_{B}T W[G;i\psi] = -\int d\bold{r}\frac{\varepsilon (\nabla \psi)^2}{2}-k_{B}T\int d\bold{r}\sum\limits_{\alpha}\bar{n}_{\alpha}(\bold{r}) +\\  k_{B}T\int d\bold{r}\int d\bold{r}' \Sigma(\bold{r},\bold{r}')\int\limits_{0}^{1} d\tau\left(G(\bold{r},\bold{r}';\tau)-G(\bold{r},\bold{r}')\right).
\end{multline}
It should be noted that we have the option to calculate the functional determinants using an alternative approach, as described in Appendix A.

For the case of point-like charges of ions, when $\varrho_{\alpha}(\bold{r})=q_{\alpha}\delta(\bold{r})$, eq. (\ref{Omega_ion_gas_}) simplifies to~\cite{wang2015theoretical,wang2016inhomogeneous}
\begin{multline}
\label{Omega_ion_gas}
\Omega = -k_{B}T W[G;i\psi] = -\int d\bold{r}\frac{\varepsilon (\nabla \psi)^2}{2}-k_{B}T\int d\bold{r}\sum\limits_{\alpha}\bar{n}_{\alpha}(\bold{r}) +\\  \int d\bold{r} \mathcal{I}(\bold{r})\int\limits_{0}^{1} d\tau\left(G(\bold{r},\bold{r};\tau)-G(\bold{r},\bold{r})\right),
\end{multline}
where
\begin{equation}
\mathcal{I}(\bold{r})=\frac{1}{2}\sum\limits_{\alpha}q_{\alpha}^2 \bar{n}_{\alpha}(\bold{r})
\end{equation}
is the local ionic strength with the local ionic concentrations
\begin{equation}
\label{chem_eq_ion}
\bar{n}_{\alpha}(\bold{r})=\frac{\delta \Omega}{\delta u_{\alpha}(\bold{r})}=z_{\alpha}A_{\alpha}(\bold{r})e^{-\beta q_{\alpha}\psi(\bold{r})-\beta u_{\alpha}(\bold{r})}.
\end{equation}

Eqs. (\ref{chem_eq_ion}) determine the chemical equilibrium conditions for ions, which can be rewritten in terms of the chemical potentials
\begin{equation}
\mu_{\alpha} = \bar{\mu}_{\alpha}+q_{\alpha}\psi+\frac{q_{\alpha}^2}{2}G(\bold{r},\bold{r}) +u_{\alpha},
\end{equation}
where the intrinsic chemical potentials of species are
\begin{equation}
\bar{\mu}_{\alpha}= k_{B}T\ln (\bar{n}_{\alpha}\Lambda_\alpha^3).
\end{equation}
Intermediate Green function $G(\bold{r},\bold{r}';\tau)$ is governed by the following equation
\begin{equation}
\label{Green_func_tau}
\left(- \varepsilon\nabla^2+\tau\Sigma\right)G(\bold{r},\bold{r}';\tau)=\delta(\bold{r}-\bold{r}').
\end{equation}

In the bulk phase with $u_{\alpha}=0$ and under thermodynamic limit where $n_{\alpha}$ becomes constant, the function $\psi(\bold{r})$ tends to zero, leading to a translation-invariant Green function, i.e.
\begin{equation}
G(\bold{r},\bold{r}')=G(\bold{r}-\bold{r}')=\int \frac{d\bold{k}}{(2\pi)^3}G(\bold{k})e^{i\bold{k}(\bold{r}-\bold{r}')}.
\end{equation}
As it follows from eqs. (\ref{Green_func}) and (\ref{Green_func_tau}),
\begin{equation}
G(\bold{k})=\frac{1}{\varepsilon(k^2+\varkappa^2(\bold{k}))},
\end{equation}
\begin{equation}
G(\bold{k};\tau)=\frac{1}{\varepsilon(k^2+\tau\varkappa^2(\bold{k}))},
\end{equation}
where we have introduced the screening function~\cite{khokhlov1982theory,borue1988statistical,lue2006variational_,budkov2019statistical}
\begin{equation}
\varkappa^2(\bold{k})=\frac{1}{k_{B}T\varepsilon}\sum\limits_{\alpha}n_{\alpha}|\varrho_{\alpha}(\bold{k})|^2
\end{equation}
with the Fourier-images of the inner charge densities (form-factors) of the ions, $\varrho_{\alpha}(\bold{k})$.

Thus, eq. (\ref{Omega_ion_gas}) leads to well known expression \cite{lue2006variational_,budkov2019statistical} for the bulk osmotic pressure $P_b=-\Omega/V$
\begin{equation}
\label{P}
P_b=k_{B}T\sum\limits_{\alpha} n_{\alpha}+\frac{k_{B}T}{2}\int \frac{d\bold{k}}{(2\pi)^3}\left(\ln\left(1+\frac{\varkappa^2(\bold{k})}{k^2}\right)-\frac{\varkappa^2(\bold{k})}{k^2+\varkappa^2(\bold{k})}\right).
\end{equation}

In the case of point-like charges, when
\begin{equation}
\varkappa^2(\bold{k})=\kappa^2=\frac{1}{k_{B}T\varepsilon}\sum\limits_{\alpha}q_{\alpha}^2n_{\alpha}
\end{equation}
eq. (\ref{P}) results in the well-known Debye-H{\"u}ckel expression~\cite{lue2006variational_}
\begin{equation}
P_{b}=k_{B}T\sum\limits_{\alpha} n_{\alpha}-\frac{k_{B}T\kappa^3}{24\pi}.
\end{equation}

\section{Extension to arbitrary reference system}
Above, we have formulated a variational field theory for Coulomb gas, i.e. when the ions have zero excluded volume. In this section, based on the paper~\cite{budkov2022modified}, referenced in the Introduction section, we formulate an extension of this variational field theory to the case of an arbitrary reference fluid system. Note that the first formulation of variational field theory, taking into account the steric interactions between ions within the lattice gas model, has been presented in~\cite{shen2018polyelectrolyte,shen2017electrostatic,agrawal2022electrostatic_}. The variational field theory of charged rigid particles in electrolyte solutions, where steric interactions are taken into account on the level of the second virial terms, has been formulated in~\cite{lue2006variational_}.

Let us assume that, apart from the Coulomb interactions, the ions interact with each other solely through repulsive potentials, denoted as $U_{\alpha\gamma}(\bold{r})$, so that the total potential energy of the system is
\begin{equation}
\label{U_2}
U=\frac{1}{2}\left(\hat{\rho} G_0\hat{\rho}\right)+\frac{1}{2}\sum\limits_{\alpha\gamma}\left(\hat{n}_{\alpha} U_{\alpha\gamma}\hat{n}_{\gamma}\right)-\frac{1}{2}\sum\limits_{\alpha}N_{\alpha}U_{\alpha\alpha}(0)+\sum\limits_{\alpha}(\hat{n}_{\alpha}u_{\alpha}),
\end{equation}
where 
\begin{equation}
\left(\hat{n}_{\alpha} U_{\alpha\gamma}\hat{n}_{\gamma}\right)=\int d\bold{r}\int d\bold{r}'\hat{n}_{\alpha}(\bold{r}) U_{\alpha\gamma}(\bold{r}-\bold{r}')\hat{n}_{\gamma}(\bold{r}'). 
\end{equation}
Employing the HS transformations, we obtain the following functional integral~\cite{budkov2022modified}:
\begin{equation}
\Xi=\int \frac{\mathcal{D}\varphi}{C_0}\exp\left[-\frac{\beta}{2}\left(\varphi G_{0}^{-1}\varphi\right)\right]\Xi_{R}[\varphi]
\end{equation}
where we have introduced the grand partition function of the reference system with pure repulsive interactions between ions
\begin{equation}
\Xi_{R}=\int \frac{\mathcal{D}\Phi}{\mathcal{N}_{U}}e^{-\frac{\beta}{2}\sum\limits_{\alpha,\gamma}(\Phi_{\alpha}U^{-1}_{\alpha\gamma}\Phi_{\gamma})+W[\Phi;\chi]}
\end{equation}
with the auxiliary functional
\begin{equation}
\label{W}
W[\Phi;\chi]=\sum\limits_{\alpha}\bar{z}_{\alpha}\int d\bold{r} e^{i\beta \Phi_{\alpha}+i\beta\chi_{\alpha}},
\end{equation}
normalization constant 
\begin{equation}
\mathcal{N}_{U}=\int {\mathcal{D}\Phi}e^{-\frac{\beta}{2}\sum\limits_{\alpha,\gamma}(\Phi_{\alpha}U^{-1}_{\alpha\gamma}\Phi_{\gamma})},
\end{equation}
and short-hand notations
\begin{equation}
\chi_{\alpha}=\varrho_{\alpha}\varphi+iu_{\alpha}.
\end{equation}
The inverse matrix operator, $U^{-1}$, is determined by the integral relation
\begin{equation}
\int d\bold{r}^{\prime\prime}\sum\limits_{\lambda} U_{\alpha\lambda}^{-1}(\bold{r}-\bold{r}^{\prime\prime})U_{\lambda\gamma}(\bold{r}^{\prime\prime}-\bold{r}^{\prime})=\delta_{\alpha\gamma}\delta(\bold{r}-\bold{r}^{\prime}).
\end{equation}

Now, as in previous section, let us apply the shift $\varphi\rightarrow \varphi + i\psi$ and transition to trial Green function $G$:
\begin{multline}
\Xi = \int \frac{\mathcal{D}\Phi}{\mathcal{N}_{U}}e^{-\frac{\beta}{2}\sum\limits_{\alpha,\gamma}(\Phi_{\alpha}U^{-1}_{\alpha\gamma}\Phi_{\gamma})}\frac{C}{C_0}\int \frac{\mathcal{D}\varphi}{C}e^{-\frac{\beta}{2}\left(\varphi G^{-1}\varphi\right)}\times\\\exp\bigg{[}-\frac{\beta}{2}\left(\varphi [G_0^{-1}-G^{-1}]\varphi\right)-i\beta (\varphi G_0^{-1}\psi)+\\\frac{\beta}{2}(\psi G_{0}^{-1}\psi)+\sum\limits_{\alpha}\bar{z}_{\alpha}\int d\bold{r} e^{i\beta \Phi_{\alpha}+i\beta \varrho_{\alpha}\varphi-\beta \varrho_{\alpha}\psi -\beta u_{\alpha}}\bigg{]}.
\end{multline}
Considering only the first cumulant over new Gaussian measure with the trial Green function $G$, we can obtain the following
\begin{multline}
\Xi\approx \exp\left[\ln\frac{C}{C_0}-\frac{1}{2}tr\left(G[G_{0}^{-1}-G^{-1}]\right)+\frac{\beta}{2}(\psi G_{0}^{-1}\psi)\right]\times\\ \int \frac{\mathcal{D}\Phi}{\mathcal{N}_{U}}\exp\left[-\frac{\beta}{2}\sum\limits_{\alpha,\gamma}(\Phi_{\alpha}U^{-1}_{\alpha\gamma}\Phi_{\gamma})+\sum\limits_{\alpha}\bar{z}_{\alpha}\int d\bold{r} A_{\alpha}(\bold{r})e^{i\beta \Phi_{\alpha}-\beta \varrho_{\alpha}\psi -\beta u_{\alpha}}\right],
\end{multline}
where $\bar{z}_{\alpha}=z_{\alpha}e^{\frac{\beta}{2}U_{\alpha\alpha}(0)}$. By employing the approach based on the cluster expansion formulated in a recent paper by one of us~\cite{budkov2022modified}, and assuming sufficiently small ranges of the potentials $U_{\alpha\gamma}(\bold{r})$, we obtain the following approximation for the grand partition function
\begin{equation}
\Xi \approx \exp W[G;\psi],
\end{equation}
where we have introduced the following auxiliary functional
\begin{equation}
W[G;\psi] = \frac{1}{2}\ln\frac{Det G}{Det G_0}-\frac{1}{2}tr\left(G[G_{0}^{-1}-G^{-1}]\right)+\frac{\beta}{2}(\psi G_{0}^{-1}\psi)+\beta \int d\bold{r}P\left(\{\bar{\mu}_{\alpha}\}\right)
\end{equation}
with the local pressure of the reference fluid system, $P$, which depends on the intrinsic chemical potentials of the ions
\begin{equation}
\bar{\mu}_{\alpha}=\mu_{\alpha}-\varrho_{\alpha}\psi-\frac{1}{2}\varrho_{\alpha} G\varrho_{\alpha}-u_{\alpha}.
\end{equation}
To ensure the self-consistency of the theory, it is necessary to determine appropriate "closures" for the functions $\psi$ and $G$. We choose these closures based on the extremum conditions for the functional $W[G;\psi]$, that is,
\begin{equation}
\frac{\delta W}{\delta \psi(\bold{r})}=0,~  \frac{\delta W}{\delta G(\bold{r},\bold{r}')}=0, 
\end{equation}
which yield
\begin{equation}
\label{Eq_Poisson_2}
\nabla^2\psi(\bold{r})=-\frac{1}{\varepsilon}\sum\limits_{\gamma}\varrho_{\gamma}\bar{n}_{\gamma}(\bold{r})
\end{equation}
and
\begin{equation}
\label{Green_func_2}
\left(- \varepsilon\nabla^2+\Sigma\right)G(\bold{r},\bold{r}')=\delta(\bold{r}-\bold{r}'),
\end{equation}
where the integral operator $\Sigma$ in this case possesses the following kernel
\begin{equation}
\label{kernel_2}
\Sigma(\bold{r},\bold{r}')=\beta \sum \limits_{\alpha}\int d\bold{x} \bar{n}_{\alpha}(\bold{x})\varrho_{\alpha}(\bold{r}-\bold{x})\varrho_{\alpha}(\bold{r}'-\bold{x}).
\end{equation}
Note that the intrinsic chemical potentials, $\bar{\mu}_{\alpha}$, can, in principle, be expressed in terms of the local concentrations using the relations 
\begin{equation}
\bar{n}_{\alpha}(\bold{r})=\frac{\delta \Omega}{\delta u_{\alpha}(\bold{r})}=\frac{\partial P}{\partial \bar{\mu}_{\alpha}(\bold{r})}.
\end{equation}

The GTP takes the form
\begin{multline}
\label{Omega}
\Omega = -k_{B}T W[G;\psi] = -\int d\bold{r}\frac{\varepsilon (\nabla \psi)^2}{2}-k_{B}T\int d\bold{r}P_0(\bold{r}) +\\  k_{B}T\int d\bold{r}\int d\bold{r}' \Sigma(\bold{r},\bold{r}')\int\limits_{0}^{1} d\tau\left(G(\bold{r},\bold{r}';\tau)-G(\bold{r},\bold{r}')\right),
\end{multline}
where $P_{0}(\bold{r})=P(\{\bar{\mu}_{\alpha}(\bold{r})\})$ is the local osmotic pressure.

It is important to note that this approach, unlike the variational method used in the previous section for the Coulomb gas, does not yield the optimal variational estimate for the GTP. However, it can be seen as a logical extension of the mean-field theory based on the saddle-point approximation~\cite{budkov2022modified}, incorporating both electrostatic correlations and steric interactions between the ions. Similarly to modified PB equations \cite{budkov2022modified}, this approach enables the use of different reference systems, such as symmetric and asymmetric lattice gas~\cite{maggs2016general_,vasileva2023theory,blossey2017structural}, as well as a hard sphere mixture model~\cite{kolesnikov2022electrosorption,huang2021hybrid}.

For the symmetric lattice gas model for which we have the following ansatz
\begin{equation}
P=\frac{k_{B}T}{v}\ln\left(1+\sum\limits_{\gamma}e^{\beta \bar{\mu}_{\gamma}}\right), 
\end{equation}
so that the local concentrations are 
\begin{equation}
\bar{n}_{\alpha}=\frac{1}{v}\frac{e^{\beta \bar{\mu}_{\alpha}}}{1+\sum\limits_{\gamma}e^{\beta \bar{\mu}_{\gamma}}}.
\end{equation}
The self-consistent field equations for this reference system have been obtained recently in ref. \cite{shen2017electrostatic,shen2018polyelectrolyte,agrawal2022electrostatic_} within a different approach. For the case of the ideal gas reference system for which $P=\sum_{\alpha}\Lambda_{\alpha}^{-3}e^{\beta \bar{\mu}_{\alpha}}$ present theory transforms into the previously discussed (see also~\cite{wang2010fluctuation}) variational field theory.

Employing the calculations similar to those in the previous section, it becomes apparent that for the case of the bulk phase of a Coulomb fluid with point-like charges of the ions, the theory with the reference system as a symmetric lattice gas leads to the following equation of state
\begin{equation}
\label{eq_of_st}
P_b=-\frac{k_{B}T}{v}\ln\left(1-v\sum\limits_{\gamma}n_{\gamma}\right)-\frac{k_{B}T\kappa^3}{24\pi},
\end{equation}
where $n_{\gamma}$ are the bulk ion concentrations, satisfying the electroneutrality condition $\sum_{\gamma}q_{\gamma}n_{\gamma}=0$. 

Although formulated above approach considers the steric interactions of the ions, it approximates the electrostatic contribution to the total osmotic pressure in the bulk phase by the Debye-H{\"u}ckel expression (see eq. (\ref{eq_of_st})).

\section{Stress tensor}
Now, we would like to discuss how to calculate the stress tensor from the derived above GTP functional. For this purpose, we apply the general covariant approach presented in our recent work~\cite{brandyshev2023noether}. In this approach, the stress tensor can be obtained using the following expression
\begin{equation}
\label{sigma_deriv_metric}
\sigma_{ik}(\bold{r})=\frac{2}{\sqrt{g(\bold{r})}}\frac{\delta \Omega}{\delta g_{ik}(\bold{r})}\bigg{|}_{g_{ik}=\delta_{ik}},
\end{equation}
where $g_{ij}$ is the metric tensor, and $g=\det{g_{ij}}$ -- its determinant.\par

We consider only the case of point-like charges of the ions. In this case, the self-consistent field equations can be written in the form 

\begin{equation}
\label{Eq_Poisson_3}
\nabla^2\psi(\bold{r})=-\frac{1}{\varepsilon}\sum\limits_{\gamma}q_{\gamma}\bar{n}_{\gamma}(\bold{r}),
\end{equation}
\begin{equation}
\label{Green_func_3}
\varepsilon\left(-\nabla^2+\varkappa^2(\bold{r})\right)G(\bold{r},\bold{r}')=\delta(\bold{r}-\bold{r}'),
\end{equation}
where 
\begin{equation}
\label{}
\varkappa^2(\bold{r})=\frac{1}{k_{B}T\varepsilon}\sum_\alpha q_\alpha^2\bar{n}_\alpha(\bold{r}).
\end{equation}

Before applying eq. (\ref{sigma_deriv_metric}), let us express the GTP, $\Omega = -k_{B}TW[G;\psi]$, in general covariant form\cite{brandyshev2023noether,brandyshev_budkov_2023}. Thus, we have
\begin{equation}\label{}
\begin{aligned}
W[G;\psi]=W_0[G;\psi]+W_1[G;\psi],
\end{aligned}
\end{equation}
where
\begin{equation}\label{}
\begin{aligned}
W_0[G;\psi]=\beta \int d\bold{r} \sqrt{g}\frac{\varepsilon(\nabla\psi)^2}{2}+ \beta \int d\bold{r} \sqrt{g}P(\{\bar{\mu}_\alpha\}),
\end{aligned}
\end{equation}\begin{equation}\label{}
\begin{aligned}
W_1[G;\psi]=\frac{1}{2}\bigg(tr \ln G-tr \ln G_0\bigg)-\frac{1}{2}tr\bigg(G\bigg[G_0^{-1}-G^{-1}\bigg]\bigg),
\end{aligned}
\end{equation}
with $\bar{\mu}_{\alpha}(\bold{r})=\mu_{\alpha}-q_{\alpha}\psi(\bold{r}) - {q_{\alpha}^2}G(\bold{r},\bold{r})/2$. Here, we consider only the case of $u_{\alpha}(\bold{r})=0$.

For composition of two integral operators,
\begin{equation}\label{}
C=AB,
\end{equation}
we have
\begin{equation}
\label{3}
C(\bold{r},\bold{r}')=\int d\bold{r}'' \sqrt{g(\bold{r}'')}A(\bold{r},\bold{r}'')B(\bold{r}'',\bold{r}').
\end{equation}
The action of the operator $A$ on a function $f(\bold{r})$ is determined by
\begin{equation}\label{4}
Af(\bold{r})=\int d\bold{r}' \sqrt{g(\bold{r}')} A(\bold{r},\bold{r}')f(\bold{r}').
\end{equation}
The trace variation is
\begin{equation}
\label{}
\delta tr(A)=\int d\bold{r} \sqrt{g(\bold{r})}\delta A(\bold{r},\bold{r})+\frac{1}{2}\int d\bold{r} \sqrt{g(\bold{r})}g^{ij}(\bold{r})\delta g_{ij}(\bold{r}) A(\bold{r},\bold{r}),
\end{equation}
which can be rewritten as
\begin{equation}\label{8}
\delta tr(A)=tr(\bar{\delta} A),
\end{equation}
where we have introduced the infinitesimal operator $\bar{\delta} A$ that has the kernel determined by the definition
\begin{equation}\label{24}
\bar{\delta} A(\bold{r},\bold{r}')=\delta A(\bold{r},\bold{r}')+\frac{1}{2} A(\bold{r},\bold{r}')g^{ij}(\bold{r}')\delta g_{ij}(\bold{r}').
\end{equation}
Thus, we have
\begin{equation}\label{}
\delta tr(A)=\int d\bold{r} \sqrt{g(\bold{r})}\bar{\delta} A(\bold{r},\bold{r}).
\end{equation}
To calculate the functional derivative with respect to the metric tensor, we take into account that
\begin{multline}
\frac{\delta W}{\delta g_{ik}(\bold{r})}=\int d\bold{x} \left(\frac{\delta W}{\delta\psi(\bold{x})}\right)_{G,g}\frac{\delta \psi(\bold x)}{\delta g_{ik}(\bold{r})}+\\\int d\bold{x}\int d\bold{x}' \left(\frac{\delta W}{\delta G(\bold{x},\bold{x}')}\right)_{\psi ,g}\frac{\delta G(\bold{x},\bold{x}')}{\delta g_{ik}(\bold{r})}+\left(\frac{\delta W}{\delta g_{ik}(\bold{r})}\right)_{\psi,G}=\left(\frac{\delta W}{\delta g_{ik}(\bold{r})}\right)_{\psi,G}.
\end{multline}
Thus, variation over the metric tensor have to be performed at constant (in variational sense) $\psi$ and $G$. Bearing in mind this circumstance, we can derive the variation of $W_0$:
\begin{equation}\label{}
\begin{aligned}
\delta W_0[G;\psi]=\beta\int d\bold{r}\sqrt{g}\delta g_{ik}\bigg(\frac{\varepsilon}{2}g^{ik}g^{mn}\partial_m\psi\partial_n\psi+g^{ik}P\bigg),
\end{aligned}
\end{equation}
where we took into account that \cite{weinberg1972gravitation,landau1975classical}
\begin{equation}\label{5}
\delta(\sqrt{g})=\frac{1}{2}\sqrt{g}g^{ij}\delta g_{ij},\quad \delta g^{ij}=-g^{im}g^{jn}\delta g_{mn}
\end{equation}
and $\partial_{i}=\partial/\partial{x}_{i}$ is the partial coordinate derivative. Note that summation over repeated coordinate indices is implied. Thus, using a determination
\begin{equation}
\sigma^{(0)}_{ik}=-\frac{2}{\beta\sqrt{g}}\frac{\delta W_0[G;\psi]}{\delta g_{ik}}\bigg|_{g_{ik}=\delta_{ik}},
\end{equation}
we can get the first part of the stress tensor
\begin{equation}
\label{}
\begin{aligned}
\sigma^{(0)}_{ik}=\varepsilon\partial_i\psi\partial_k\psi-
\frac{\varepsilon}{2}\delta_{ik}\partial_l\psi\partial_l\psi-P\delta_{ik}.
\end{aligned}
\end{equation}
Note that the latter expression coincides with the stress tensor in the mean-field approximation obtained in ref.~\cite{budkov2022modified}

The variation of the second part of the functional is
\begin{equation}\label{}
\begin{aligned}
\delta W_1[G;\psi]=-\frac{1}{2}tr\bigg(G_0^{-1}\bar{\delta}G_0\bigg)-\frac{1}{2}tr\bigg(G\bar{\delta}G_0^{-1}\bigg).
\end{aligned}
\end{equation}
Therefore, we use the relations
\begin{equation}\label{}
\bar{\delta}G_0^{-1}G_0+G_0^{-1}\bar{\delta}G_0=0,
\end{equation}
\begin{equation}\label{}
G_0^{-1}(\bold{r}',\bold{r})=-\varepsilon\Delta\delta(\bold{r}-\bold{r}'),\quad \bar{\delta}G_0^{-1}(\bold{r}',\bold{r})=-\varepsilon\Delta'\delta(\bold{r}-\bold{r}'),
\end{equation}
where $\Delta = \nabla^2$ is the Laplacian and $\Delta'$ is the differential operator, determined by
\begin{equation}\label{}
\begin{aligned}
\Delta'f=Df-
\frac{1}{2}g^{mn}\delta g_{mn}\Delta f.
\end{aligned}
\end{equation}
Thus, we can show that the variation is
\begin{equation}\label{}
\begin{aligned}
\delta W_1[G;\psi]=\frac{\varepsilon}{2}tr(\Delta'\bar{G}),
\end{aligned}
\end{equation}
where
\begin{equation}
\bar{G}=G-G_0,
\end{equation}
and the special differential operator
\begin{equation}\label{}
\begin{aligned}
Df=\frac{1}{2\sqrt{g}}\partial_{i}\bigg(\sqrt{g}g^{mn}\delta g_{mn}g^{ij}\partial_{j}f\bigg)
-\frac{1}{\sqrt{g}}\partial_{i}\bigg(\sqrt{g}\delta g_{mn}g^{im}g^{jn}\partial_{j}f\bigg)
\end{aligned}
\end{equation}
has been introduced.

After the same calculations as in \cite{brandyshev_budkov_2023}, we obtain
\begin{equation}\label{}
\begin{aligned}
\sigma^{(1)}_{ij}(\bold{r})=\frac{\varepsilon k_{B}T}{2}\lim_{\bold{r}'\rightarrow \bold{r}}\hat{D}_{ij}\bar{G}(\bold{r},\bold{r}'),
\end{aligned}
\end{equation}
where the following differential operator
\begin{equation}
\hat{D}_{ij}=\delta_{ij}\partial_k\partial_k +\delta_{ij}\partial_k\partial_k'\\
-\partial_i\partial_j'-\partial_j\partial_i'
\end{equation}
has been introduced.

Now, it is necessary to prove that the tensor
\begin{equation}
\sigma_{ik}=\sigma^{(0)}_{ik}+\sigma^{(1)}_{ik}
\end{equation}
is the stress tensor. For this purpose, we have to show that its divergence is equal to zero, i.e. $\partial_{i}\sigma_{ik}=0$.

Using eqs. (\ref{Eq_Poisson_3}) and (\ref{Green_func_3}), we obtain the following
\begin{equation}\label{34}
\begin{aligned}
\partial_{i}\sigma^{(1)}_{ik}(\bold{r})=\frac{\varepsilon k_{B}T}{2}G(\bold{r},\bold{r})\partial_{k}\varkappa^2(\bold{r}).
\end{aligned}
\end{equation}
Using the same equations, we obtain the following results
\begin{equation}
\partial_{i}\sigma^{(0)}_{ik}(\bold{r}) =\frac{\varepsilon k_{B}T}{2} \varkappa^2(\bold{r}) \partial_{k} G(\bold{r},\bold{r}).
\end{equation}
Thus, we get the following
\begin{equation}
\label{eq_cond_1}
\partial_{i}\sigma_{ik} = \frac{\varepsilon k_{B}T}{2}  \partial_{k} (\varkappa^2(\bold{r})  G(\bold{r},\bold{r})).
\end{equation}
It can be observed that the approximate variational functional $\Omega=-k_{B}TW[G;\psi]$ does not yield a divergenceless stress tensor. This is because the adopted approximation does not guarantee consistency with the mechanical equilibrium, which must be derived from the exact GTP functional. However, it is surprising that equation (\ref{eq_cond_1}) can still be reformulated in the form of a conservation law
\begin{equation}
\partial_{i}T_{ik}=0,
\end{equation}
where
\begin{equation}
T_{ik}=\sigma_{ik}-\frac{\varepsilon k_{B}T}{2} \varkappa^2(\bold{r})  G(\bold{r},\bold{r}) \delta_{ik}.
\end{equation}
Bearing in mind that with the use of the self-consistent field equations (\ref{Eq_Poisson_3}), (\ref{Green_func_3})
\begin{equation}
\sigma^{(1)}_{ik}(\bold{r})=\frac{\varepsilon}{2}\bigg(k_{B}T\varkappa^2(\bold{r})G(\bold{r},\bold{r})
+\mathcal{D}_{ll}(\bold{r})\bigg)\delta_{ik}-\varepsilon \mathcal{D}_{ik}(\bold{r}),
\end{equation}
the tensor $T_{ik}$ can be eventually presented as
\begin{equation}
\label{Tij}
T_{ik}=-P\left(\{\bar{\mu}_{\alpha}\}\right)\delta_{ik}+\varepsilon\left(\partial_i\psi\partial_k\psi-
\frac{1}{2}\delta_{ik}\partial_l\psi\partial_l\psi\right)+\varepsilon\left(\frac{1}{2}\mathcal{D}_{ll}(\bold{r})\delta_{ik}- \mathcal{D}_{ik}(\bold{r})\right),
\end{equation}
where
\begin{equation}
\mathcal{D}_{ik}(\bold{r})= k_{B}T\lim_{\bold{r}'\rightarrow \bold{r}}
\partial_i\partial_k' G(\bold{r},\bold{r}')
\end{equation}
is the autocorrelation function of the electric field fluctuations. We removed the terms that include the bare Green function, $G_{0}(\bold{r},\bold{r}')$, since they pertain to identically divergenceless tensor and do not contribute to mechanical forces~\cite{brandyshev_budkov_2023}. In practical calculations, it is necessary to subtract from $G(\bold{r},\bold{r}')$ the bare Green function, $G_{0}(\bold{r}-\bold{r}')$, for the infinite space case (for example, see Appendix B). The divergenceless tensor $T_{ik}$ can be interpreted as the total stress tensor, which is consistent with the self-consistent field equations (\ref{Eq_Poisson_3}) and (\ref{Green_func_3}). In eq. (\ref{Tij}) the first term in the right-hand side represents the hydrostatic isotropic stress tensor; the second term represents the standard Maxwell stress tensor; the third term represents the contribution of fluctuations in the local electric field around the mean-field configuration. {Eq. (\ref{Tij}) is the main new result of this paper. The latter allows us, in principle, to calculate the force of interaction between conductive bodies with electrified surfaces immersed in an ionic fluid by the following expression 
\begin{equation}
F_{i}= \oint T_{ik}n_{k}d\mathcal{A},
\end{equation}
where the integration is performed over the surface of one of the bodies; $d\mathcal{A}$ is the elementary area of the body's surface, $n_{k}$ are the components of the external normal to the body at a certain point. In the case that there are no specific interactions between a conductive body and ions, we can simplify the analysis by replacing the need to integrate over the body's surface with a more general integration over an arbitrary closed surface surrounding the conductive body.}

In the presence of the external fields, i.e. when $\bar{\mu}_{\alpha}(\bold{r})=\mu_{\alpha}-q_{\alpha}\psi(\bold{r}) - {q_{\alpha}^2}G(\bold{r},\bold{r})/2 - u_{\alpha}(\bold{r})$, tensor (\ref{Tij}) yields
\begin{equation}
\label{eq_cond}
\partial_{i}T_{ik}-\sum\limits_{\alpha}\bar{n}_{\alpha}\partial_{k}u_{\alpha}=0.
\end{equation}
The latter equation represents the mechanical equilibrium condition of the Coulomb fluid in the presence of external fields. We will use this equation below to obtain the expression for the disjoining pressure. Eq. (\ref{eq_cond}) is nothing more but the hydrostatic equation for the Coulomb fluids.

{The force acting on the conductive surface immersed in the ionic fluid in presence of specific interactions between body's surface and ions can be calculated by
\begin{equation}
F_{i}=\oint T_{ik}n_{k}d\mathcal{A}+\int f_{i} dV,
\end{equation}
where the first integral is taken over an arbitrary closed surface around the macroscopic body, whereas the second integral -- over the volume between surface of integration and the surface of the macroscopic body; $f_{i}=-\sum_{\alpha}\bar{n}_{\alpha}{\partial u_{\alpha}}/{\partial x_{i}}$ are the components of the total nonelectrostatic volume density force describing the aforementioned specific interactions.}

\section{Disjoining pressure}

Let us consider the case of a Coulomb fluid that is confined in a slit-like pore with infinite charged walls. We assume that the walls create external potentials 
\begin{equation}
u_{\alpha}(z)=\phi_{\alpha}(z)+\phi_{\alpha}(H-z),
\end{equation}
where $\phi_{\alpha}(z)$ is the single-wall external potential of the ion of the $\alpha$th kind at point with coordinate $z\in [0,H]$. The identical walls are located at $z=0$ and $z=H$. The disjoining pressure is determined by
\begin{equation}
\Pi = -\frac{\partial (\Omega/\mathcal{A})}{\partial H}-P_{b},
\end{equation}
where $\mathcal{A}$ is the total area of the walls. Taking into account that the functional $\Omega = -k_{B}T W[G;\psi]$ achieves the extremum at functions that satisfy the self-consistent field equations (\ref{Eq_Poisson_3}) and (\ref{Green_func_3}), we obtain the following.
\begin{equation}
\Pi = -\sum\limits_{\alpha}\int\limits_{-\infty}^{\infty}dz\bar{n}_{\alpha}(z)\phi^{\prime}_{\alpha}(z)-P_{b},
\end{equation}
where we used the relations $\bar{n}_{\alpha}(H-z)=\bar{n}_{\alpha}(z)$, stemming from the symmetry determined by the identity of the walls. Integrating the mechanical equilibrium condition,
\begin{equation}
\label{eq_cond_2}
\frac{d T_{zz}}{dz}- \sum\limits_{\alpha}\bar{n}_{\alpha}(z)u_{\alpha}'(z)=0,
\end{equation}
from $z=H/2$ to $z=\infty$ and taking into account that $T_{zz}(\infty)=0$, we obtain
\begin{equation}
-T_{zz}(H/2)=\sum\limits_{\alpha}\int\limits_{H/2}^{\infty} d z \bar{n}_{\alpha}(z)u_{\alpha}'(z)=\sum\limits_{\alpha}\int\limits_{H/2}^{\infty} d z \bar{n}_{\alpha}(z)\phi_{\alpha}'(z)-\sum\limits_{\alpha}\int\limits_{-\infty}^{H/2} d z \bar{n}_{\alpha}(z)\phi_{\alpha}'(z).
\end{equation}
Therefore,
\begin{multline}
\label{Pi2}
\Pi=-\sum\limits_{\alpha}\int\limits_{-\infty}^{\infty}dz\bar{n}_{\alpha}(z)\phi^{\prime}_{\alpha}(z)-P_{b}=
-\sum\limits_{\alpha}\int\limits_{-\infty}^{H/2}dz\bar{n}_{\alpha}(z)\phi^{\prime}_{\alpha}(z)-\sum\limits_{\alpha}\int\limits_{H/2}^{\infty}dz\bar{n}_{\alpha}(z)\phi^{\prime}_{\alpha}(z)-P_b=\\-T_{zz}(H/2)-P_b-2\sum\limits_{\alpha}\int\limits_{H/2}^{\infty}dz\bar{n}_{\alpha}(z)\phi_{\alpha}^{\prime}(z).
\end{multline}
Thus, taking into account that $\bar{n}_{\alpha}(z)=0$ at $z \geq H$ (impermeable wall), we eventually obtain the following result
\begin{equation}
\label{disj_press}
\Pi=P_n-P_b-2\sum\limits_{\alpha}\int\limits_{H/2}^{H}dz\bar{n}_{\alpha}(z)\phi_{\alpha}^{\prime}(z),
\end{equation}
where $P_{n}=-T_{zz}(H/2)$ is the normal pressure at the middle of the pore. If the pore is sufficiently large and the range of the wall potential is sufficiently small, we can neglect the integral in the right-hand side of eq. (\ref{disj_press}). 
Note that the same expression for the disjoining pressure has been obtained within the pure mean-field theory in ref.~\cite{budkov2022modified}.

Now, we would like to specify the normal stress at the pore midpoint. For slit-like pore the Green function can be presented as
\begin{equation}\label{}
G(\bold{r},\bold{r}')=G(\bm{\rho}-\bm{\rho}'|z,z'),
\end{equation}
where $\bm{\rho}$ is the two-dimensional vector lying in the plane of the pore and the $z$-axis is perpendicular to the pore wall. Let us consider the two-dimensional Fourier transform \cite{wang2015theoretical,wang2016inhomogeneous,agrawal2022electrostatic_}
\begin{equation}\label{}
G(\bm{\rho}-\bm{\rho}'|z,z')=\int \frac{d^2\bold{q}}{(2\pi)^2}e^{-i\bold{q}(\bm{\rho}-\bm{\rho}')}G(q|z,z'),
\end{equation}
where $G(q|z,z')$ is the even function of $\bold{q}$ depending only on the vector modulus $q=|\bold{q}|$.
The cross elements of the tensor $\mathcal{D}_{ik}(z)$ are
\begin{equation}\label{}
\mathcal{D}_{xy}(z)=\mathcal{D}_{xz}(z)=\mathcal{D}_{yz}(z)=0,
\end{equation}
whereas its diagonal elements are
\begin{equation}\label{}
\mathcal{D}_{xx}(z)=\int \frac{d^2\bold{q}}{(2\pi)^2}q_x^2Q(q,z),~\mathcal{D}_{yy}(z)=\int \frac{d^2\bold{q}}{(2\pi)^2}q_y^2Q(q,z),
\end{equation}
\begin{equation}\label{}
\mathcal{D}_{zz}(z)=\int \frac{d^2\bold{q}}{(2\pi)^2}\mathcal{D}_{zz}(q,z),
\end{equation}
where
\begin{equation}
\mathcal{D}_{zz}(q,z)=k_{B}T\lim_{z'\rightarrow z}\partial_z\partial_{z'}G(q|z,z'),
\end{equation}
\begin{equation}
Q(q,z)=k_{B}T\lim_{z'\rightarrow z}G(q|z,z'),
\end{equation}
so that 
\begin{equation}
\mathcal{D}_{ll}(z)=\int \frac{d^2\bold{q}}{(2\pi)^2}\bigg(q^2Q(q,z)+\mathcal{D}_{zz}(q,z)\bigg).
\end{equation}
The Fourier-image of the Green function, $G(q|z,z')$, can be found from the equation
\begin{equation}\label{}
\varepsilon \bigg(-\partial^2_z+q^2+\varkappa^2(z)\bigg)G(q|z,z')=\delta(z-z').
\end{equation}
Therefore, the normal pressure in eq. (\ref{disj_press}) is
\begin{equation}
\label{Pn}
P_{n}=P_{m}+\varepsilon \left(\mathcal{D}_{zz}\left(\frac{H}{2}\right)-\frac{1}{2}\mathcal{D}_{ll}\left(\frac{H}{2}\right)\right),
\end{equation}
where $P_{m}=P_0(H/2)$ is the osmotic pressure of the ions at the pore middle and we have used that $\psi'(H/2)=0$.

To conclude this section, we would like to estimate the asymptotic of the second term in the right-hand side of eq. (\ref{Pn}) corresponding to the electric field fluctuations at $H\to \infty$. Taking into account that in this asymptotic $\varkappa(z)\simeq\kappa$ and that~\cite{agrawal2022electrostatic_}
\begin{equation}
G(q|z,z')\simeq\frac{e^{-\sqrt{q^2+\kappa^2}|z-z'|}}{2\varepsilon \sqrt{q^2+\kappa^2}},~G_0(q|z,z')=\frac{e^{-q|z-z'|}}{2\varepsilon q},
\end{equation}
we obtain
\begin{multline}
\varepsilon \left(\mathcal{D}_{zz}\left(\frac{H}{2}\right)-\frac{1}{2}\mathcal{D}_{ll}\left(\frac{H}{2}\right)\right)\simeq \\\frac{k_{B}T}{8\pi}\int\limits_{0}^{\infty}dq q\left(2q-\sqrt{q^2+\kappa^2}-\frac{q^2}{\sqrt{q^2+\kappa^2}}\right)=-\frac{k_B T \kappa^3}{24\pi},
\end{multline}
i.e., as it should be, Debye-H{\"u}ckel expression.

{We can also calculate the asymptotic behavior of the disjoining pressure between two conductive non-electrified walls (within the pore volume $\psi=0$) at $\kappa H\gg 1$. The results has the form (see Appendix B)
\begin{equation}
\label{Picor}
\Pi=-\frac{I}{2\pi \varepsilon H}\exp\left[-2\kappa H\right],
\end{equation}
where $I=\sum\limits_{\alpha}q_{\alpha}^2n_{\alpha}/2$ is the bulk ionic strength. We notice that the exponentially damped contribution creates an effective attractive interaction between the walls at large distances, resulting from the electrostatic correlations of ions~\cite{levin2002electrostatic}.}

\section{Discussion}
{As explained above, the contribution of electrostatics to the bulk osmotic pressure is described by the Debye-H{\"u}ckel expression, which is expressed in terms of the Debye screening length, $r_{D}=\kappa^{-1}$ (see equation (\ref{eq_of_st})). However, if we consider the self-consistent field equation (\ref{Eq_Poisson_3}) for the electrostatic potential, we can derive another screening length in the linear approximation (weak electrostatics), which depends on the choice of the reference system (see also refs.~\cite{maggs2016general,kumari2022nature}). This discrepancy arises when we approximate the GTP by using the local osmotic pressure, which depends on the chemical potentials, which in turn include the self-energies~\cite{wang2010fluctuation,wang2013effects,wang2014continuous} of the ions. As a result, the current form of variational field theory predicts the weak electrostatic coupling limit for the bulk osmotic pressure. However, a recent study~\cite{agrawal2023ion} has shown that in the case of multivalent ions, a symmetric lattice gas reference system can qualitatively describe the attractive force between like-charged dielectric walls in an electrolyte solution at small interwall distances.}

{We also note that the formulation of the theory does not treat dielectric discontinuities and thus can be applied to the modeling of electric double layers at the metal-electrolyte interfaces. However, theory can be directly generalized to take into account the dielectric heterogeneity (dependence of dielectric permittivity on coordinates) on interfaces with such macroscopic dielectrics, as membranes~\cite{buyukdagli2019like,buyukdagli2021contribution,blossey2023poisson}. Comprehensive examples of such variational field theories that account for dielectric heterogeneity can be found in papers~\cite{wang2013effects,wang2014continuous,wang2015theoretical,wang2016inhomogeneous}. We also note that the present theory does not take into account the static polarizability and dipole moment of the ions~\cite{budkov2021theory,budkov2020two,frydel2011polarizable} and polar molecules of solvent~\cite{abrashkin2007dipolar,coalson1996statistical,iglivc2010excluded}. They can be taken into account in the same manner as in recently presented pure mean-field theories (see Models II and III as classified in \cite{budkov2022modified}).}

\section{Conclusions}
{In conclusion, within the variational field theory framework~\cite{wang2010fluctuation}, taking into account the electrostatic correlations of the ions, we have extended our previous mean-field formalism~\cite{budkov2022modified}. Using previously formulated general covariant approach~\cite{brandyshev_budkov_2023} we have derived a total stress tensor that considers the electrostatic correlations of ions. This is accomplished through an additional term that depends on the autocorrelation function of the local electric field fluctuations. It is important to note that this stress tensor has the potential to be used in calculating the force of interaction between conductive bodies that have electrified surfaces and are immersed in an ionic fluid. Utilizing the derived total stress tensor and applying the mechanical equilibrium condition, we have obtained a general expression for the disjoining pressure of the Coulomb fluids, confined in a pore with a slit-like geometry. Using this equation, we have derived an asymptotic expression for the disjoining pressure in a slit-like pore with non-electrified conductive walls. The general expression obtained for the disjoining pressure allows us to directly calculate the disjoining pressure of a Coulomb fluid within a slit-like pore with conductive walls. This can be achieved by numerically solving the self-consistent field equations, eliminating the need for computationally intensive differentiation of the grand thermodynamic potential with respect to the pore thickness. Our present theory serves as the foundation for future modeling of the mechanical stresses that occur in electrode pores with conductive charged walls immersed in liquid phase electrolytes, extending beyond mean-field theory. The focus of our future publications will involve performing the corresponding numerical calculations for physically relevant reference fluid systems, such as hard sphere mixtures and asymmetric lattice gases.}

\textbf{Data availability statement.} {\sl The data supporting the findings of this study are available in the article.}

{\bf Acknowledgements.}  
This work is an output of a research project implemented as part of the Basic Research Program at the National Research University Higher School of Economics (HSE University). We would like to express our gratitude to the anonymous reviewers for their valuable comments, which have helped us make significant improvements to the text.

\appendix
\section{An alternative method of calculations of the logarithm of the fraction of functional determinants.}
In this appendix, we demonstrate an alternative approach to calculate the logarithm of the fraction of functional determinants, different from the charging method\cite{wang2015theoretical,lue2006variational_}. Therefore, we can establish a chain of equalities to illustrate this method:
\begin{multline}
\label{DetG}
\ln \frac{Det G}{Det G_0}=tr\ln{G}-tr\ln{G_0}=\sum\limits_{n}\ln\frac{v_n}{u_n}=\int\limits_{0}^{\infty}ds
\sum\limits_{n}\left(\frac{1}{s+u_{n}}-\frac{1}{s+v_{n}}\right)\\
=\int\limits d\bold{r}\int\limits_{0}^{\infty}ds \sum\limits_{n}\left(\frac{1}{s+u_{n}}\psi_{n}(\bold{r})\psi_{n}(\bold{r})-\frac{1}{s+v_{n}}\phi_n(\bold{r})\phi_n(\bold{r})\right)\\
=\int d\bold{r}\int\limits_{0}^{\infty}ds\left(\left(s+G^{-1}\right)^{-1}-\left(s+G_0^{-1}\right)^{-1}\right)\delta(\bold{r}-\bold{r}')\biggr\rvert_{\bold{r}=\bold{r}'}\\
=\int d\bold{r}\int\limits_{0}^{\infty}ds\left(\mathcal{G}(\bold{r},\bold{r}|s)-\mathcal{G}_0(\bold{r},\bold{r}|s)\right)=
\int\limits_{0}^{\infty}ds \left(tr \mathcal{G}(s)-tr \mathcal{G}_0(s) \right),
\end{multline}
where we have introduced the auxiliary Green functions, $\mathcal{G}$ and $\mathcal{G}_{0}$, determined by the following equations
\begin{equation}
(G^{-1}+s)\mathcal{G}(\bold{r},\bold{r}'|s)=\delta(\bold{r}-\bold{r}')
\end{equation}
and
\begin{equation}
(G_0^{-1}+s)\mathcal{G}_0(\bold{r},\bold{r}'|s)=\delta(\bold{r}-\bold{r}')
\end{equation}
the eigenvalues $u_{n}$ and $v_n$ of the operators $G^{-1}$ and $G_0^{-1}$, respectively, and the corresponding orthonormal eigenfunctions ${\psi_n(\bold{r})}$ and ${\phi_n(\bold{r})}$. We have also used the functional completeness conditions for the eigenfunctions, that is
\begin{equation}
\sum\limits_{n}\phi_n(\bold{r})\phi_n(\bold{r}')=\sum\limits_{n}\psi_n(\bold{r})\psi_n(\bold{r}')=\delta(\bold{r}-\bold{r}').
\end{equation}

\section{Derivation of eq. (\ref{Picor}).}

{In this appendix, we give a derivation of eq. (\ref{Picor}). Let us consider the case of non-electrified conductive surface, so that within the pore volume $\psi(z)=0$. We neglect also the specific interactions of the ions with the walls ($u_{\alpha}=0$). The disjoining pressure can be obtained by
\begin{equation}
\label{Pi2}
\Pi = P_{n}-P_{b}=P_{m}+\varepsilon \left(\mathcal{D}_{zz}\left(\frac{H}{2}\right)-\frac{1}{2}\mathcal{D}_{ll}\left(\frac{H}{2}\right)\right)-P_{b}.
\end{equation}
Let us estimate the asymptotic behavior of the disjoining pressure at $H\to \infty$, where $\varkappa(z)\simeq \kappa$, so the the Green function can be obtained from the equation
\begin{equation}
\label{}
\varepsilon \bigg(-\partial^2_z+q^2+\kappa^2\bigg)G(q|z,z')=\delta(z-z')
\end{equation}
with the following boundary conditions
\begin{equation}
G(q|0,z')=G(q|H,z')=0,
\end{equation}
\begin{equation}
~\varepsilon\left(\frac{\partial G(q|z'-0,z')}{\partial z}-\frac{\partial G(q|z'+0,z')}{\partial z}\right)=1,~ G(q|z'-0,z')=G(q|z'+0,z').
\end{equation}}

The solution is
\begin{equation}
\label{G}
G(q|z,z')=
\begin{cases}
\frac{\sinh \kappa_{q}z\left(\cosh \kappa_q z'-\coth\kappa_q H\sinh \kappa_q z'\right)}{\varepsilon \kappa_q}, &z<z',\\
\frac{\sinh \kappa_{q}z'\left(\cosh \kappa_q z-\coth\kappa_q H\sinh \kappa_q z\right)}{\varepsilon \kappa_q}, &z>z',
\end{cases}
\end{equation}
where $\kappa_q=\sqrt{q^2+\kappa^2}$.
Thus, we have
\begin{multline}
\label{corr}
{\varepsilon} \left(\mathcal{D}_{zz}\left(\frac{H}{2}\right)-\frac{1}{2}\mathcal{D}_{ll}\left(\frac{H}{2}\right)\right)=\frac{k_{B}T}{4}\int \frac{d^2\bold{q}}{(2\pi)^2} \left(2q-\frac{\kappa_q}{\tanh \frac{\kappa_q H}{2}}-\frac{q^2\tanh \frac{\kappa_q H}{2}}{\kappa_q}\right)\\
=\frac{k_{B}T}{4}\int\frac{d^2\bold{q}}{(2\pi)^2}\left(2q-\sqrt{q^2+\kappa^2}-\frac{q^2}{\sqrt{q^2+\kappa^2}}\right)\\
-\frac{k_{B}T}{4}\int\frac{d^2\bold{q}}{(2\pi)^2}\left(\kappa_q\left(\coth\frac{\kappa_q H}{2} -1\right)+\frac{q^2}{\kappa_q}\left(\tanh\frac{\kappa_q H}{2} -1\right)\right).
\end{multline}
Furthermore, 
\begin{equation}
\label{Pn3}
P_{n}\simeq P_{0}+\delta P_n,
\end{equation}
where $P_0$ is the bulk pressure of the reference system (without electrostatic interactions) and $\delta P_n$ -- its perturbation, caused by the electrostatic correlations of the ions
\begin{equation}
\label{Pcor}
\delta P_n=-\frac{1}{2}\sum\limits_{\alpha}q_{\alpha}^2 n_{\alpha}\int \frac{d\bold{q}}{(2\pi)^2}\delta G\left(q\bigg{|}\frac{H}{2},\frac{H}{2}\right).
\end{equation}
We have also introduced the variable
\begin{equation}
\delta G\left(q\bigg{|}\frac{H}{2},\frac{H}{2}\right)=G\left(q\bigg{|}\frac{H}{2},\frac{H}{2}\right)-\lim\limits_{H\to \infty}G\left(q\bigg{|}\frac{H}{2},\frac{H}{2}\right)=\frac{\tanh\frac{\kappa_q H}{2}-1}{2\varepsilon \kappa_q}
\end{equation}
and took into account that $n_{\alpha}=\partial{P}/\partial{\mu_{\alpha}}$. Therefore, taking into account (\ref{corr}) and that
\begin{equation}
P_{b}=P_0 -\frac{k_{B}T\kappa^3}{24\pi},
\end{equation} 
and substituting (\ref{Pn3}) into equation (\ref{Pi2}), we obtain
\begin{equation}
\Pi = -\frac{k_{B}T}{2}\int\frac{d^2\bold{q}}{(2\pi)^2}\kappa_q\left(\coth \kappa_{q}H-1\right)=-\frac{k_{B}T}{2\pi}\int\limits_{0}^{\infty}dq\,q\frac{\kappa_q}{e^{2\kappa_q H}-1}.
\end{equation}

The latter yields
\begin{equation}
\label{disj_press_asymp}
\Pi\simeq
\begin{cases}
-\frac{k_B T}{8\pi H^3}, & \kappa H\ll 1\,\\
-\frac{k_B T\kappa^2}{4\pi H}e^{-2\kappa H},&\kappa H\gg 1.
\end{cases}
\end{equation}

The first regime refers to the classical limit of van der Waals interaction between metals (refer to equation (4.30) in paper \cite{dzyaloshinskii1961general}). The second regime corresponds to eq. (\ref{Picor}) in the main text, which has been obtained for the first time in ref. \cite{ninham1997ion} by different method.

\bibliographystyle{aipnum4-2}
\bibliography{name}

\end{document}